# Government Spending and Money Supply Roles in Alleviating Poverty in Africa

[1]Gbatsoron Anjande, [2]Simeon T. Asom, [3]Ngutsav Ayila, [4]Bridget Ngodoo Mile and [5]Victor Ushahemba Ijirshar

[1,2,3,4,5]Department of Economics, Benue State University, Makurdi-Nigeria.
[1]_ganjande@yahoo.com_ [2]_asomsimeon59@gmail.com_ [3]_ngutsavayila@gmail.com_
[4]_milengodoo@gmail.com_ [5]_ijirsharvictor@gmail.com_

*Abstract*: This study examines the roles of government spending and money supply on alleviating poverty in Africa. The study used 48 Sub-Saharan Africa countries from 2001 to 2017. The study employed one step and two-step system GMM and found that both the procedures have similar results. Different specifications were employed and the model selected was robust, with valid instruments and absence of autocorrelation at the second order. The study revealed that government spending and foreign direct investment have significant negative influence on reducing poverty while money supply has positive influence on the level of poverty in the region. The implication of the finding is that monetary policy tool of money supply has no strong influence in combating the menace of poverty. The study therefore recommends that emphasis should be placed on increasing more of government spending that would impact on the quality of life of the people in the region through multiplier effect, improving the financial system for effective monetary policy and attracting foreign direct inflows through enabling business environment in Africa.

## 1.0 Introduction

Poverty is seen as a multidimensional problem that include: social, political, and cultural issues, among others. Poverty has been a global problem and it affects different people in different continents, regions and countries in different ways and magnitude. Binuyo (2014) simply asserts that there is no country or region that is immune from poverty, however, the magnitude varies from region to region and country to country.

Africa is a resource endowed region. However, most of the Africans are poor (Edrees, Azali, Hassan & Nor, 2015). Poverty in Africa is therefore a serious challenge facing the region or continent. This include lack of human needs such as: food, access to health care, access to education, safe drinking water and electricity. Statistics have shown that about 76% of the poorest countries in the world (such as: Liberia, Somalia, Ethiopia and Zimbabwe) are located in Africa (Edrees, Azali, Hassan & Nor, 2015). Although global poverty has been on the decline except in some countries in Sub-Saharan Africa (Oriavwote & Ukawe, 2018). According to World Bank (2016), the number of people living in extreme poverty in Sub-Saharan Africa has grown substantially since 1990. Poverty level in Africa rose even in the last 20 years (Adigan, Awoyemi & Omonona, 2011; Ravallion & Chen, 2004). It has become an endemic disease in developing countries especially in Sub-Saharan Africa countries.

As a result of the negative consequences of poverty, governments from Sub-Saharan Africa countries have embarked on several fiscal and monetary policies to eradicate or combat poverty in the region. Government spending as one of the fiscal policy tools plays an important role in poverty reduction. According to Keynes theory, government spending may increase the aggregate demand which further stimulates the economic growth, employment and reduces poverty (Mehmood & Sadiq, 2010). Similarly, Farayibi and Owuru (2016) argues that fiscal policy administration through the mechanism of government spending plays an important role in poverty reduction, increase per capita income and finally culminates into economic growth and development. Therefore, government spending can affect economic growth positively and cause reduction in poverty level through the provision of social goods/services and infrastructural facilities needed for sustainable growth. Among the government strategies is the World Bank's Poverty Reduction Strategies Programme (PRSP) that emphasized investment in health, primary education and social sector in order to boast human capital formation for reducing poverty (Asghar, Hussain & Rehman, 2012).

More so, there is growing concern about the distributive impacts of money supply as a monetary policy tool. However, poverty and social exclusion therefore serves as two key indicators of a government's commitment. In the light of the above, this study is set to measure the roles of government spending and money supply in alleviating poverty in Africa. This study considers 48 African countries from the Sub-Saharan region within the 21$^{st}$ century (2001 to 2017).

## 2.0   Literature Review

### 2.1   Conceptual Clarification

**Money Supply:** Jhingan (2004) defined money supply as the total amount of money in the economy at a point in time. This simply means that money supply is the total stock of monetary media of exchange available to a society for use in connection with the economic activity of the country. It includes: cash, coins, and money held in savings and checking accounts for short-term payments and investments. The money supply reflects the extent of liquidity that different money instruments have on an economy. Based on the size and type of account in which a liquid instrument belongs, money supply is broadly classified into: $M_1$, $M_2$, and $M_3$. The narrow money supply ($M_1$) is defined as currency outside bank plus demand deposits of commercial banks plus domestic deposits with the Central Banks less Federal Government deposits at commercial banks (Ifionu & Akinpelumi, 2015). In simple terms, $M_1$ is defined as:

$$M_1 = C + D \qquad (1)$$

Where: $M_1$ = Narrow money supply, $C$ = Currency outside banks, $D$ = Demand deposits. $M_2$ is generally defined as broad money. Ajayi (1979) noted that $M_2$ includes not only notes and coins and bank current accounts, but also 7-days bank deposits and some building society deposits. Thus, broad money supply ($M_2$) is defined as $M_1$ plus quasi money. Quasi-money is seen as the sum of savings and time deposits with commercial banks. $M_2$ is defined as:

$$M_2 = C + D + T + S \tag{2}$$

Where: $M_2$ = Board money, $T$ = Time deposit, $S$ = Savings deposits, $C$ = Currency outside bank and $D$ = Demand deposits. In this study, we considered broad money growth which is define as the sum of currency outside banks; demand deposits other than those of the central government; the time, savings, and foreign currency deposits of resident sectors other than the central government; bank and traveler's checks; and other securities such as certificates of deposit and commercial paper (World Bank, 2017).

**Poverty:** Poverty can be regarded as the state of not being able to meet the basic needs (Gideon & Thaddeaus, 2016). These include: shelter, clothing, and food, among others. Ajakaiye and Adeyeye (2001) have broadly conceptualized poverty in four ways. These include: lack of access to basic needs/goods; a result of lack or impaired access to productive resources; outcome of inefficient use of common resources; and result of exclusion mechanism. Poverty is lack of basic needs/goods, essentially economic or consumption oriented. Basic goods are nutrition, shelter/housing, water, healthcare, and access to productive resources including education, working skills, tools, political and civil rights to participate in decisions concerning socio-economic conditions (Adeyeye, 1999). The first three are the basic needs/goods necessary for survival. Impaired access to productive resources (agricultural land, physical capital and financial assets) leads to absolute low income, unemployment, undernourishment, among others.

Laderchi, Saith and Stewart (2003) defined poverty into four approaches: monetary, capabilities, social exclusion and participation. They considered monetary approach as the most common approach. It measures the shortfall in income or consumption below poverty line. The capabilities approach is the lack of some basic capabilities to function. The functioning can vary from such elementary physical ones as being well nourished, being adequately clothed and sheltered, avoiding preventable morbidity, among others, to more complex social achievements such as taking part in the life of the community, being able to appear in public without shame and so on. The opportunity of converting personal income into capabilities to function depends on a variety of personal circumstances (including age, gender, and disabilities, among others) and social surroundings (including epidemiological characteristics, physical and social environments, public services of health and education and so on). The definition extends poverty analysis to a wider scope than the monetary approach. It also focuses on the individuals and how they can realize their potentials. Social exclusion is the state of being wholly or partially excluded from full participation in the society in which the subjects live. The fourth approach dwells on participation. It is the only approach that has attempted to include the poor in defining their own poverty. This study adopts the definition by World Bank (2017) that poverty is or National poverty headcount ratio is the percentage of the population living below the national poverty lines. National estimates are based on population-weighted subgroup estimates from household surveys.

**Government Spending**: The term government spending is used most times interchangeably with government expenditure. It plays key role in the operation of every economy (Keynes, 1936). It refers to expenses incurred by the government for the maintenance of itself and provision of public goods, services and works needed to foster or spur economic growth and improve the welfare of people in the society. Government spendings are generally categorized into: expenditures on administration, defense, internal securities, health, education, foreign affairs, among others and have both capital and recurrent components (Aigheyisi, 2013).

The capital expenditure refers to the amount spent in the acquisition of fixed (productive) assets (whose useful life extends beyond the accounting or fiscal year), as well as expenditure incurred in the upgrade/improvement of existing fixed assets such as lands, building, roads, machines and equipment, and so on, including intangible assets. It also include: expenses on capital projects such as: roads, airports, health, education, electricity generation, and so on. Capital expenses are usually aimed at increasing the assets of a state and they give rise to recurrent expenditure. Expenditure in research also falls within this component of government spending. Capital expenditure is usually seen as expenditure creating future benefits, as there could be some lags between when it is incurred and when it takes effect on the economy. This includes: capital expenditures on administration, economic services and social community services.

Recurrent expenditure on the other hand refers to expenditure on purchase of goods and services, wages and salaries, operations as well as current grants and subsidies (usually classified as transfer payments). Recurrent expenditure, excluding transfer payments, is also referred to as government final consumption expenditure (Aigheyisi, 2013). Generally, recurrent expenditure refers to government expenses on administration, security, maintenance of public goods, interest payment on loans, etc. It includes expenditures on general administration, defence, internal security, national assembly, education, health, other social and community services, agriculture, construction, transport & communication, other economic services, public debt servicing, pensions and gratuities, contingencies/subventions and other/other CFR charges (CBN, 2017).

Okoro (2013) defines government expenditure as the value of goods and services provided through the public sector. It can also be described as the expenses incurred by the government in the provision of public goods and services. Thus, government expenditure is an important instrument for government in controlling the economy. It is the main instrument used by governments especially in developing countries to promote economic growth, income per capita income, reduce unemployment and poverty which are essential ingredients for sustainable development. Economic growth brings about a better standard of living of the people through provision of better infrastructure, health, housing, education services and improvement in agricultural productivity and food security as earlier noted. Almost all the sectors of developing economies demand more budgetary allocations every fiscal year. For instance, the agricultural sector under the Maputo Declaration of 2003 requires African governments to increase expenditure on agricultural sector to at least 10 percent of the national budgetary resources (New Partnership for Africa's Development-NEPAD, 2014). According to Keynes (1936), government spending is a major component of national income as seen in the expenditure approach to measuring national income as: $Y = C+I+G+(X-M)$ where, Y is the national income, C is the consumption expenditure, G is the government spending, X is the export component and M is the import component. This implies that government spending is a key determinant of the size of the economy and generally economic growth. However, it could act as a two-edged sword: It could significantly boost aggregate output, especially in developing countries where there are massive market failures and poverty traps, and it could also have adverse consequences such as unintended inflation and boom-bust cycles. The effectiveness of government spending in expanding economy and fostering rapid economic growth depends on whether it is productive or unproductive. All things being equal, productive government expenditure would have positive effect on the economy, while unproductive expenditure would have the reverse effect.

## 2.2 Theoretical Review

In the early part of the twentieth century, Fisher formulated the quantity equation which was a part of the classical theory (Nikitin, 1995). The classical theory states that when the quantity of money in circulation rises, real money supply reduces leading to a rise in general price level (Ireland, 2014). Fisher in 1911 provided a concise equation to capture the classical theory of money supply as follows:

$$MV = PT \qquad 3$$

Where M=stock of money; V=velocity of money; P=price level; and T=transactions. Denoting the national output by Y and replacing transactions national output since the former is not measurable unlike the measurable. Hence, substituting it into equation 1.1, it becomes:

$$MV = PY \qquad 4$$

The equation (1.2) explains the quantity theory of money. The theory posits that if velocity of money and national output are assumed constant, money supply is transmitted into a rise in the price level (Dornbusch, Fischer & Richard, 2008). As Nikitin (1995) observed the quantity theory of money was flawed on several grounds, which inter alia, includes the constancy assumptions of V and Y, as well as the strict proportionality condition between M and P. The neo-classical economists failed to provide any remedy to the great depression of 1930s. This led Keynes theory of money.

Keynes rejected the idea of an unfailing self-regulatory economic system enshrined in the neoclassical doctrine that became grossly disappointing during the great depression. Keynes suggested that an increase in government spending or effective demand paves way out of depression. Thus, the theory of demand-pull inflation draws from the fact that when there is a positive output gap at full employment, an inflationary gap will rise. Keynes ideas can be demonstrated in the IS-LM framework where demand for output determines supply up to the full employment level. Keynes emphasize the idea of government intervention.

Monetarists further refuted the Keynesian prescription of government interventions in economic affairs including the idea and practice of deficit spending (Nikitin, 1995). They hinged on laissez faire belief that an economy is always near or close to full employment (thus, negative output gap is negligible) and that changes in the quantity of money in circulation will only affect the price level and output in the long-run.

In the classical theory of poverty, the market is self-adjusting or regulating and resources are efficiently assigned to production units. Thus, poverty is seen not as result of market failure but poor economic decisions of individuals such as being lazy or being uneducated (Omari & Muturi, 2016; Davis & Sanchez-Martinez, 2014). It may also be attributed to inefficient allocation of productive resources. This is because, Omari and Muturi (2016) argues that living in deprivation is as a result of individual decisions and that better choices and hard work are sufficient to lift one out of poverty. In the words of Davis and Sanchez-Martinez (2014) as viewed by non-poor is that people who live in poverty deserves it and they tend to choose and nurture a culture of poverty which leads to intergenerational poverty (Davis & Sanchez-Martinez, 2014). According to Omari and Muturi (2016), government interventions to eradicate poverty of this nature through government spending are highly discouraged as it interferes with automatic market mechanism and may result to inefficiency ultimately. This spending can however be channeled to providing support activities and programmes that would facilitate individuals to engage in productive activities to earn a living wage/eking a living. The classical emphasized that government intervention to stem poverty is seen to rather reinforce poverty as it makes individuals to be dependent on the welfare thereby serving as disincentives to individuals' efforts to towards being more productive.

In the views of Keynes, he criticized the culture of poverty proposed in classical economics not to be consistent among the poor. Keynes argued that poverty is caused by economic underdevelopment and lack of human capital (Omari & Muturi, 2016; Jung & Smith, 2006). Issues of market failures such as uncertainties can cause critical economic situation given that the poor are more vulnerable to shocks that affect their income. The theory argues that the poor are impoverished due to external factors mostly beyond their control. Hence, government intervention is seen as a means to promoting economic development and welfare (Omari & Muturi, 2016; Davis

& Sanchez-Martinez, 2014). For instance, during the great depression of 1930s, Keynes argued that government intervention through expansionary fiscal policies or increased government spending was necessary to stimulate aggregate demand and create jobs thereby reducing unemployment and poverty. Thus, government intervention stimulates the growth of an economy and via multiplies effect reduce poverty.

The study adopts the Keynesian theory who asserts that increases in government expenditure leads to high aggregate demand and rapid growth in national income (Keynes, 1936). This has capacity of reducing poverty among the populace. He favored government intervention to correct market failures and criticize the classical economists (Keynes, 1936). Keynes advocated a countercyclical fiscal policy in which, during the boom periods, the government ought to cut expenditure, and during periods of economic recession, government requires to undertake deficit spending. Keynes categorized government spending as an exogenous variable that can generate economic growth instead of an endogenous phenomenon. He believed the role of the government to be crucial as it can avoid depression by increasing aggregate demand and thus, switching on the economy again by the multiplier effect. It is a tool that bring stability in the short-run but this need to be done cautiously as too much of public spending lead to inflationary situations while too little of it leads to poverty and unemployment (Keynes, 1936).

According to Keynes (1936), the state of an economy is determined by four parameters: the money supply, the demand functions for consumption (or equivalently for savings) and for liquidity, and the schedule of the marginal efficiency of capital determined by the existing quantity of equipment and the state of long-term expectation (Keynes, 1936).

## 2.3 Money Supply and Poverty Reduction

This study examined the relationship between money supply or generally monetary policy and economic growth. Several authors have also attempted the examination of the influence of monetary policy on economic growth using either panel data or time series data. Goshit and Longduut (2016) examined the effectiveness of indirect monetary policy instruments in reducing poverty in Nigeria using multiple regression covering from 1986 to 2012. Ordinary Least Squares (OLS) technique was employed and the findings showed that money supply significantly reduce poverty in Nigeria. In a similar vein, Ekobena (2014) also investigated the influence of monetary policy on inequality and poverty using panel system GMM estimation technique. The study covered United States and Economic and Monetary Community of Central Africa (EMCCA1) countries from 1986 to 2011. Findings from the study revealed that monetary policy and particularly interest rate and poverty are positively related in the United States, implying that increase in the rate of interest causes high level of poverty. However, for EMCCA countries, conventional monetary policy does not affect income distribution and poverty but rather through the quantitative easing channel. However, the study focused on the effectiveness of other monetary policy instruments such as interest rate and quantitative easing other than conventional monetary policy of money supply. In terms of the impact of macroeconomic policies, Amjad and Kemal (1997) assessed the impact of macroeconomic policies on poverty using multiple regression covering 1984-85 to 1990-91. The study found that policies pursued under the Structural Adjustment Programme have tended to increase the poverty levels due to decline in growth rates, withdrawal of subsidies on agricultural inputs and consumption, decline in employment, increase in indirect taxes, among others while a strategy for poverty eradication, the employment programmes as well as promotion of informal sector enterprises are imperative.

The ultimate objective of monetary policy is to control the volume of money in circulation in such a way as to promote sound economic performance and high living standards of the citizens. This gives the citizens confidence in the currency as a store of value, unit of account and medium of exchange, so that they can make sound economic and financial decisions. Money supply is one among the monetary policy tools. Monetary policy generally impacts on the wellbeing of individuals depending on the policy measures put in place (Gideon & Thaddeaus, 2016). For instance, money supply affects poverty by influencing the cost and availability of credit at commercial banks. An expansionary monetary policy therefore reduces the cost of credit and thus, boosts investments that generates employment and in turn reduces poverty and increases wellbeing (CBN, 2011).

Romer and Romer (1989) have identified at least five channels through which monetary policy can affect long-run income distribution. First, the redistribution caused by swings in unanticipated inflation directly raise inequality. Second, the reduction in physical capital investment caused by uncertainty and financial markets disruptions raise the average return on capital and depress wages; thus widened the income distribution. Third, offsetting this, inflation may shift the burden of taxation away from labour towards capital. Fourth, the markets cause by inflation and macroeconomic instability reduce not just physical investment, but human capital investment. This thwarts an important mechanism by which inequality can be mitigated. Finally, inflation and macroeconomic volatility may harm some sectors of the economy disproportionately. Goshit (2014) has opined that monetary policy can also influence poverty reduction through the attainment of high level of employment in the economy. When the objective of monetary policy is to achieve high level of employment in an economy, money supply assumes an expansionary dimension. First, an expansion in money supply leads directly to increased expenditure on goods and services and in turn increases employment to produce the extra goods and services being demanded. This increased employment invariably enhances income and thereby reduce income poverty in an economy. On the other hand, Ajayi (1979) opines that the alternative view in the transmission mechanism in which the increased money supply is seen to have led to a fall in interest rates, which in turn increases investment expenditure and thus, increased employment could be more potent in poverty reduction. Therefore, improving the quality and quantity of employment opportunities links economic growth to poverty reduction. Hence, a development strategy that fully employs a country's human resources and raises the returns to labour becomes an effective instrument for reducing poverty. It is on this premise that Romer and Romer (1989) submitted that an alternative monetary policy which focuses on real variables including employment and faster GDP growth is both feasible and necessary if any developing country is to make more rapid progress in reducing poverty and generating sustainable development. It is on this backdrop that Oni (2006) stressed that to generate employment opportunities, the monetary policy must encourage employment-generating investment, facilitate sustainable economic expansion and maintain macroeconomic stability. This study examines the influence of government spending and money supply on poverty in Africa.

## 2.4    Government Spending and Poverty Reduction

This study has examined the relationship between government spending and economic growth. Hence, previous empirical studies on the influence of government spending on economic growth were reviewed.

Chude, Chude, Anah and Chukwunulu (2019) examined the relationship between government expenditure, economic growth and poverty reduction in Nigeria covering 1980 to 2013. Using unit root tests, bound test co-integration approach and error correction techniques within an ARDL framework which yields more robust estimates, it was found that government spending affect economic growth positively and significantly. Findings emerged from this study also showed that government expenditure has significant short run impact on poverty reduction in Nigeria. Similarly, Oriavwote and Ukawe, (2018) investigated the impact of government expenditure on poverty reduction in Nigeria covering 1980 and 2016. The study used OLS and causality test and found that government expenditure on education has a significant positive impact on per capita income. The study also found bidirectional relationship between government expenditure on education and per capita income in Nigeria. Disaggregating government expenditure based on sectors, Omodero and Omodero (2019) examined the impact of government expenditure by sector on poverty from 2000 to 2017. The study employed ordinary least squares technique and the result indicates that government expenditure on building and construction, agriculture, education and health had no significant impact on poverty alleviation in Nigeria which was attributed to the insufficiency of the government spending on these key sectors of the economy. Bright (2016) also assessed the effect of government expenditure on poverty incidence for Ghana covering 1960 to 2013 using the Johansen test (JH), Vector Error Correction (VECM) test and the Ordinary Least Square (OLS). The study found that poverty incidence positively correlated with government expenditure implying that poverty is not reducing with increase in government expenditure. The contrary view can be attributed to the non-stationary estimation of poverty variable using ordinary least squares technique.

Other studies such as: Rashid and Sara (2010) examined the relationship between fiscal deficits on poverty in Pakistan from 1976 to 2010. The study found a negative relationship between government expenditure and poverty and found that there exist short run and long run relationship between poverty and government expenditure. In a similar vein, Enyim (2013) examined the relationship between government spending and poverty reduction in Nigeria from 1980 to 2009 using Ordinary Least Square (OLS) technique. The study found that government spending has significant negative impact on poverty in Nigeria. Omari and Muturi (2016) also examined the impact of government expenditure by sectors on poverty level in Kenya covering 1964 to 2010. The study used Vector Autoregressive model, Co-integration analysis and error correction mechanism after ascertaining the presence of co-integration. The study found that agriculture sector and health sector expenditures have a positive and significant effect on poverty level while infrastructure sector expenditure has a negative and significant effect on poverty level. However, the effect of education sector expenditure on poverty level was not significant. Dahmardeh and Tabar (2013) assessed the relationship between government spending and poverty in Sistan and Baluchestan Province of Iran. The study examined the effects of government spending on poverty reduction from 1978 to 2008 using of Autoregressive Distributed Lag (ARDL) method. The study also found significant influence on reducing poverty in the country. As shown in the results, constructive expenditures have positive effect on poverty reduction Sistan and Baluchestan Province of Iran. Fan, Hazell and Thorat (1998) in another dimension examined the causes of the decline in rural poverty in India and quantify the effectiveness of government expenditures on poverty reduction covering 1970 to 1993. The study used simultaneous equation system. The study found that government spending has significant impact on poverty reduction.

Using panel data, Anderson, d'Orey, Duvendack and Esposito (2018) also examined the relationship between government spending and income poverty on low and middle income countries using 19 countries. Employing regression, the study found that the relationship between government spending and poverty is negative for countries in Sub-Saharan Africa though less negative than Eastern European countries and Central Asian countries. Shahrier and Lian (2018) also assessed the impact of monetary and fiscal policies on poverty line and income distribution of the bottom 20 percentile to the top 20 percentile using Financial Computable General Equilibrium (FCGE) model. The study found that expansionary fiscal policy or increase in government spending is more effective than the expansionary monetary policy in narrowing the income distribution and improving the income of the bottom 20% of the population in the short-run while monetary policy that aimed at low inflation and stable aggregate demand would permanently improve poverty incidence in the long-run.

Thus, increase in government spending is also likely to cause a rise in aggregate demand. This could also lead to higher growth and help to reduce levels of inequality and improve wellbeing. However, if an economy is close to full capacity, higher government expenditure could lead to crowding out effect (a situation when government spending rather causes reduction in private sector spending). Since, Africa is a developing region, there is dear need for increasing its expenditure. Hence, the impact of government spending or expenditure depends on the state of the economy. For instance, higher government expenditure may cause inflationary pressures and little increase in real GDP in an economy that is close to full capacity while government borrowings from the private sector acts as expansionary fiscal policy to boost economic growth when an economy is in recession. Increases in government expenditures may create a multiplier effect and further cause reduction in unemployment and poverty level and increase in aggregate demand.

The size of government spending and its effect on poverty reduction, and vice versa, has been an issue of sustained interest for decades. The relationship between government spending and poverty reduction has continued to generate series of debate among scholars as noted above. Government performs two functions- protection (and security) and provisions of some public goods. The creation of rule of law and enforcement of property rights are under protection functions. This helps to minimize risks of criminality, protect life and property, and the nation from external aggressions. Expenditure on goods such as: defense, roads, education, health, and power are under provisions of public goods. Some scholars argue that increase in government spending on socio-economic and physical infrastructures encourages economic growth and reduces poverty (Enyim, 2013; Osundina, Ebere & Osundina, 2014). These include: expenditure on infrastructure such as roads, communications and power. Sameti & Karami (2004) noted that government spending can have direct and indirect effects on poverty. The direct effects arise in the form of benefits the poor receive from expenditures on employment and welfare programs. The indirect effects arise when government investments in rural infrastructure, agricultural research, and the health and education of rural people, stimulate agricultural and nonagricultural growth, leading to greater employment and income earning opportunities for the poor, and to cheaper food. This study employs system of equation within the framework of panel system GMM to assess effects of government spending and money supply on poverty in Africa. Several studies examined the country specific impact of government spending on poverty and found that increased spending reduces poverty (Chude, Chude, Anah & Chukwunulu, 2019; d'Orey, Duvendack & Esposito, 2018; Oriavwote & Ukawe, 2018; Edrees, Azali, Hassan & Nor, 2015; Asghar, Hussain & Rehman, 2012; Rashid & Sara, 2010)

The introduction of SAP in Africa has made poverty to be arguably the most pressing economic problem of our time because rising inequality in levels of income has led to greater poverty (Goshit, 2014). Yet, the reduction of poverty is the most difficult challenge facing any developing economy just like Africa where the average of the population is considered poor and evidence shows that the number of those that are poor have continue to increase. Ogwumike (2001) estimated that more than 70 per cent of Africans live in poverty. Despite several government's policies and programmes towards poverty reduction in the economy, more than half of the population remain poor (Ogwumike, 2001). The consequences of extreme poverty in Africa are manifested in diverse ways including vices such as arm robbery, hunger, malnutrition, prostitution, illiteracy, kidnapping, oil pipes vandalism, ethno-religious and political crises and others which are inimical to economic growth and development. At the same time, fiscal policy through government spending and monetary policy through money supply are among the traditional tools can help curb poverty in an economy.

**3.0    Methodology**

**3.1    Method of Data Analysis**

This study used GMM. The study used two approaches of deciding whether difference GMM or system GMM is most preferred. This include: Blundell-Bond (1998) and Bond, Hoeffler and Temple (2001). Blundell-Bond (1998) argues that difference GMM estimator may yield both a biased and inefficient estimate of the lagged dependent variable if the parameter estimate is persistent and close to being a random walk in finite samples and particularly acute when T is short. Blundell and Bond (1998) therefore attributed the poor performance of the difference GMM estimator in such cases to the use of poor instruments. Hence, the use of system GMM estimator overcome this problem. The study also estimated the autoregressive model using pooled OLS and fixed effects approach. The pooled OLS estimate for the lagged variable of poverty was considered as an upper-bound estimate, while the corresponding fixed effects estimate was considered a lower bound estimate. The decision rule is that if the difference GMM estimate obtained is close to or below the fixed effects estimate, the difference GMM is downward biased because of weak instrumentation and a system GMM is more preferred.

**3.2    Data Needs and Sources**

All the data for the estimations were collected from World Development Indicators from 2001 to 2017. These include data on: poverty, government spending, broad money supply, foreign direct investment, domestic capital investment, gross national income per capita and overall economic freedom. These include: Angola, Benin, Botswana, Burkina-Faso, Burundi, Cameroon, Cape Verde, Central African Republic, Chad, Comoros, Democratic Republic of the Congo, Djibouti,

Equatorial Guinea, Eritrea, Ethiopia, Gabon, Gambia, Ghana, Guinea, Guinea-Bissau, Ivory Coast, Kenya, Lesotho, Liberia, Madagascar, Malawi, Mali, Mauritania, Mauritius, Mozambique, Namibia, Niger, Nigeria, Republic of the Congo, Rwanda, Sao Tome and Principe, Senegal, Seychelles, Sierra Leone, Somalia, South Africa, Sudan, Swaziland, Tanzania, Togo, Uganda, Zambia and Zimbabwe.

### 3.3 Theoretical Model

Keynes (1936) argues that increasing government spending in times of economic downturn corrects the economy back to equilibrium. Following the Keynesian model that income or poverty level (POV) of an economy depends on the level of government intervention through increased government spending (G), it can be stated as:

$$POV = f(G) \qquad 5$$

According to Keynes (1936), government intervenes in order to increase the level of aggregate demand or income. Thus, high levels of income in an economy may help in reducing poverty. This can be expressed as:

$$POV = f(G, Y) \qquad 6$$

Where GNI is the annual income of an economy. Keynes (1936) further argued that an increase in money supply ($M_2$) affects income and in turn reduces poverty.

$$POV = f(G, Y, M_2) \qquad 7$$

Keynes (1936) also argued that income is affected by investment which can either be Foreign Investment (FI) or Domestic Investment (DI) and the degree of openness (OP). The equation 7 can be restated as:

$$POV = f(G, Y, M_2, FI, DI, OP) \qquad 8$$

### 3.4 Model Specification

Using the theoretical model as stated in equation 8, with the theoretical expectation that increase in government spending, money supply, domestic capital investment, foreign direct investment inflows, gross national income and economic freedom are expected to have negative influence on poverty. Based on the above theoretical basis, the dynamic panel model for the study is stated as:

$$POV_{it} = \beta_0 + \delta POV_{i,t-1} + \beta_1 GSP_{it} + \beta_2 BMS_{it} + \beta_3 FDI_{it} + \beta_4 CIN_{it} + \beta_5 GNI_{it} + \beta_6 EFD_{it} + \varphi_i + \varepsilon_{it} \qquad 9$$

Where, POV = Poverty, GSP = Government Spending, BMS = Broad Money Supply, FDI = Foreign Direct Investment, CIN = Domestic Capital Investment, GNI = Gross National Income per capita and EFD = Overall Economic Freedom. $\beta_0$ = Intercept, $\beta_1 - \beta_6$ = Parameter coefficients to be estimated, $\varphi_i$ = Individual Specific Effect or Fixed Effect and $\varepsilon_{it}$ = An idiosyncratic error.

### 3.5    Variable Definitions

Poverty: The study used poverty headcount ratio at $1.90 a day and it is the percentage of the population living on less than $1.90 a day at 2011 international prices (World Bank, 2019). Data for this variable is sourced from The World Bank

Government spending as percent of GDP: General government final consumption expenditure (formerly general government consumption) includes all government current expenditures for purchases of goods and services (including compensation of employees) (World Bank, 2019). Data for this variable is sourced from The World Bank

Broad money supply is the sum of currency outside banks; demand deposits other than those of the central government; the time, savings, and foreign currency deposits of resident sectors other than the central government; bank and traveler's checks; and other securities such as certificates of deposit and commercial paper (World Bank, 2019). This data is sourced from The World Bank

Foreign Direct Investment, percent of GDP: Foreign direct investment are the net inflows of investment to acquire a lasting management interest (10 percent or more of voting stock) in an enterprise operating in an economy other than that of the investor (World Bank, 2019). Data for this variable is sourced from The World Bank

Capital investment as percent of GDP: Gross capital formation (formerly gross domestic investment) consists of outlays on additions to the fixed assets of the economy plus net changes in the level of inventories (World Bank, 2019). Data for this variable is sourced from The World Bank.

GNI per capita: GNI per capita is gross national income divided by midyear population (World Bank, 2019). In this study, annual percentage growth rate of GNI per capita based on constant local currency was used. Data for this variable is sourced from The World Bank.

Economic freedom, overall index (0-100): The Overall index of economic freedom has ten components grouped into four broad categories: Rule of Law; Limited Government; Regulatory Efficiency and Open Markets. The overall economic freedom is scored on a scale of 0 to 100, where 100 represents the maximum freedom. Data for this variable is sourced from The Heritage Foundation

## 4.0 Results and Discussions

### 4.1 Decision of using either difference or system GMM

The study estimated the pooled OLS, Fixed Effect (FE), difference GMM and system GMM to obtain the estimate (δ) of the lagged dependent variable ($pov_{t-1}$). The δ estimate in pooled OLS is considered biased upwards while δ estimate in Fixed Effect is considered biased downwards. As a decision rule, if the δ estimate lies below or close to FE estimate, it is biased downwards, hence, system GMM estimator will be considered, otherwise, difference GMM will be considered. The estimated result of the pooled OLS, FE, difference GMM and system GMM for δ (lagged dependent variable) are presented in Table 1

**Table 1: Decision of the use of difference and system GMM**

| Estimators | Coefficients |
|---|---|
| Pooled OLS | 1.500746 |
| Fixed Effects | 0.994518 |
| One-Step Difference GMM | 1.139221 |
| Two-Step Difference GMM | 1.102021 |
| One-Step system GMM | 1.328566 |
| Two-Step system GMM | 1.102021 |

Source: Author's computation from STATA 15 output.

From the result in Table 1, the estimate of δ in one-step and two-step difference GMM are closer to the estimate of δ in fixed effects, thus, the little difference between the estimate of δ in one-step/two-step difference GMM and fixed effects suggests that there may be much benefits to using the system GMM estimator as argued by Bond, Hoeffler and Temple (2001).

**4.2    Correlation Results**

The result of correlation analysis is presented in Table 2.

Table 2: Correlation Test Results

|     | POV | GSP | BMS | FDI | CIN | GNI | EFD |
|-----|-----|-----|-----|-----|-----|-----|-----|
| POV | 1 | | | | | | |
| GSP | -0.1186 | 1 | | | | | |
| BMS | -0.3642 | 0.3492 | 1 | | | | |
| FDI | 0.0137 | 0.0728 | 0.0073 | 1 | | | |
| CIN | -0.2572 | 0.2418 | 0.092 | 0.3463 | 1 | | |
| GNI | -0.5908 | 0.2174 | 0.3816 | 0.0109 | 0.2456 | 1 | |
| EFD | -0.2849 | 0.0844 | 0.3765 | -0.0953 | 0.1606 | 0.2789 | 1 |

Source: Authors' Computation from STATA 15 Output.

From the results of correlation test in Table 2, it implies that all the regressors are not linearly dependent on one another or exact. Hence, there is absence of multicollinearity in the model.

**4.3    The Roles of Government Spending and Money Supply on Poverty in Africa**

The estimated results of one-step and two- step system GMM for the 48 Sub-Saharan African countries in analysing the effectiveness of government spending and money supply are presented in Table 3.

**Table 3: The Result of One-step and Two-step System GMM**

| Variables | One Step System GMM POV | Two Step System GMM POV |
|---|---|---|
| L.POV | 1.329***[0.000] | 1.321***[0.000] |
|  | (-0.0597) | (-0.0771) |
| GSP | -0.0682[0.022] | -0.0443**[0.048] |
|  | (-0.024) | (-0.017) |
| BMS | 0.0572**[0.042] | 0.0621**[0.047] |
|  | (-0.0274) | (-0.0234) |
| FDI | -0.0897**[0.012] | -0.0854**[0.026] |
|  | (-0.0341) | (-0.0286) |
| CIN | 0.0896[0.265] | 0.0812[0.507] |
|  | (-0.0792) | (-0.121) |
| GNI | 0.00147***[0.000] | 0.00154***[0.001] |
|  | (-0.00034) | (-0.00042) |
| EFD | 0.0594[0.455] | 0.0103[0.914] |
|  | (-0.0788) | (-0.0948) |
| Constant | -22.17***[0.001] | -16.99**[0.033] |
|  | (-6.02) | (-7.691) |
| AR(2) | 0.15[0.883] | 0.20[0.841] |
| Hansen | 26.33[0.015] | 26.33[0.015] |
| Observations | 567 | 567 |
| Number of Country (Groups) | 44 | 44 |
| Number of Instruments | 38 | 38 |
| F-statistics | 170.27[0.000] | 109.74[0.000] |

Source: Authors' Computation from STATA 15 Output.

Note: Robust standard errors in parentheses ( ) and probability in brackets [ ]

*** p<0.01, ** p<0.05, * p<0.1

The result from Table 3 shows significant positive estimate of the lagged dependent variable ( $pov_{t-1}$ ) for the one-step and two-step system GMM estimates. The estimates of one-step system GMM are similar with two-step system GMM procedure. Hence, the study considers the estimates of two-step system GMM for statistical inferences in this study. The result shows 38 numbers of instruments and 44 number of groups. This means that the instruments in the model are not over-bloated. The result also reveals a statistically significant influence of lagged values of the dependent variable on current level of poverty in Sub-Saharan Africa. This implies that poverty has a cycle and people who live in the cycle of poverty have lagged effect of the incidence. The implication is that past levels of poverty strongly influence the current level of poverty in the region.

The result also shows that government spending has an estimated negative coefficient of –0.0442743 with the probability value of 0.048<0.05. The estimated coefficient is theoretically plausible and statistically significant at 5% significance level. The relationship indicates that increasing government spending reduces the level of poverty in sub-Saharan Africa. This implies that expansionary fiscal policy in terms of increased government spending has multiplier effect on the level of poverty in Africa.

The result also reveals that money supply has positive influence on the level of poverty. The estimated coefficient is positive (0.0621243) and statistically significant at 5% level of significance. The relationship, though not theoretically plausible, implies that increased money supply exert positive influence on the levels of poverty in Sub-Saharan Africa. The implication is that increase in money supply fuels up poverty in the region.

The result also reveals negative relationship between foreign direct investment and poverty in Sub-Saharan Africa. The estimated coefficient of foreign direct investment (-0.0853582) with the probability value of 0.026<0.05 is also statistically significant at 5% significance level. This means that foreign direct investment inflows to Africa increases investment in the region that improves the income of the host communities. The result conforms to the theoretical position of eclectic theory by Dunning (1979). He argues that there must be some kind of location advantage in the place where the company is located.

The estimated coefficients of domestic capital investment, gross national income and overall economic freedom are 0.0811689, 0.0015375 and 0.103152. Apart from the estimated coefficient of gross national income that is statistically significant at 1% significance level, the estimated coefficients of domestic capital investment and overall economic freedom are not statistically significant even at 10% significance level. The strong positive relationship of gross national income with poverty implies that higher income for Sub-Saharan African countries are skewed or titled towards very few rich persons, hence, poverty increase.

From the result, the numbers of instruments (38) were less than the number of groups (44) and there is joint significance of the variables incorporated in the model (that is, F-statistics of 109.74 and Prob. value of 0.000<0.05). The model also shows the Hansen statistic of 26.33 with its probability of 0.015<0.05. This implies that there are valid instruments in the model. The AR (2) statistic is 0.20 with probability value of 0.841. This means that the model does not suffer from second order autocorrelation. The estimated lagged dependent variable was also statistically significant at 1% significance level.

### 4.3 Year Dummies Control for Time Variations of the Dependent Variable

The trend of the year dummies control for time variations of the dependent variable across panels is depicted in Figure 1. The interpretation of the year dummies is relative to the base year, 2001.

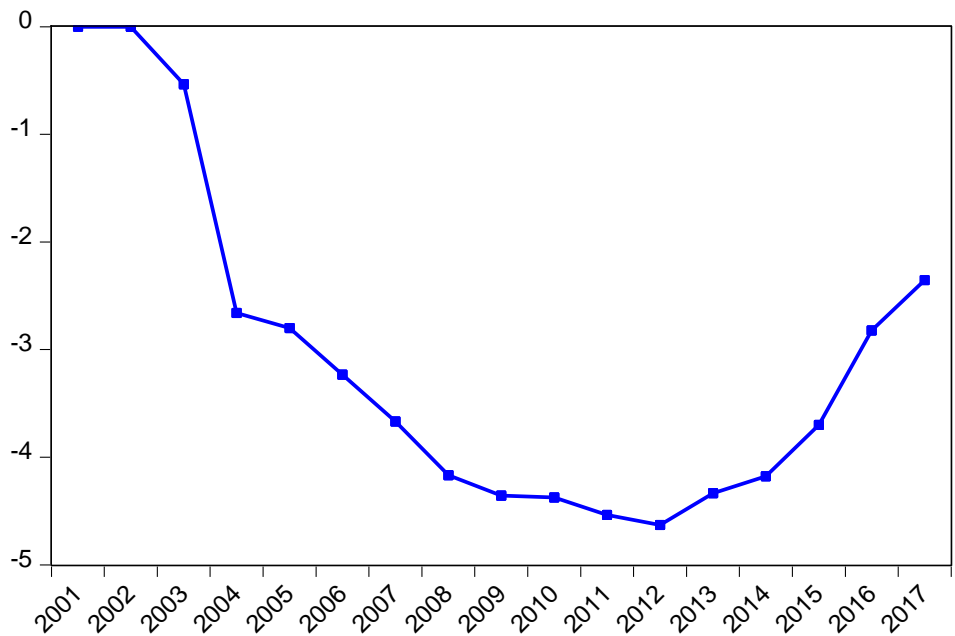

Figure 1: Poverty in Africa

The trend in Figure 1 shows that poverty in 2003 is on average and *ceteris paribus*, 53% lower than the base year 2001. The relative percentage of poverty to base year declined over time to 2012 and exhibited upward trending pattern from 2013 to 2017.

### 4.4 Summary of Findings

The study found that government spending and foreign direct investment had negative and significant influence on poverty among the African countries at 5% level of significance while money supply has positive influence on the level of poverty in the region implying that monetary policy tool of money supply has not been effective in combating the menace of poverty.

## 5.0 Conclusion/Recommendations

The study measures the impact of government spending and money supply on alleviating poverty in Sub-Saharan Africa within the 21$^{st}$ century. The study concludes that government spending has strong influence on reducing the level of poverty in Africa while money supply has not significant impact in combating the menace of poverty in the region. The study also conclude that government spending and foreign direct investment have been significant determinant of improved welfare among people in the region unlike money supply. The implication of the finding is that monetary policy tool of money supply is not significant or momentous in combating the menace of poverty in the region. This may be attributed to the level of financial development in the region as majority of the population are unbanked. Hence, implementing monetary policy of money supply may not exert strong influence on combating poverty in the region.

The study therefore recommends that African governments should increase the level of their spending especially on key productive sectors that could impact on the quality of life of the people in the region. This can be done by increasing the proportion of capital budget on yearly basis in order to achieve an improved productive economy that could affect the lives of people positively. More so, the financial system of the African countries should be improved for policy complementarity that could exert stronger influence on poverty in the region. This can be done by increasing the number of the banked population as this would help in increasing the effectiveness of monetary policy.

The study also recommends creating of enabling environment that could attract foreign direct investment inflows in the African countries. This can be done by advancing the ease of doing business and provisions of insecurity free environment and infrastructures for smooth running of businesses.